\documentstyle[amssymb,multicol,epsfig,pre,aps]{revtex}

\begin{document}

\title{Yield conditions for deformation of amorphous polymer glasses}

\author{J\"org Rottler and Mark O.~Robbins}

\address{Department of Physics and Astronomy, The Johns Hopkins
University, 3400 N.~Charles Street, Baltimore, MD 21218}

\date{\today}

\maketitle
\begin{abstract}
Shear yielding of glassy polymers is usually described in terms of the
pressure-dependent Tresca or von Mises yield criteria.  We test these
criteria against molecular dynamics simulations of deformation in
amorphous polymer glasses under triaxial loading conditions that are
difficult to realize in experiments.  Difficulties and ambiguities in
extending several standard definitions of the yield point to triaxial
loads are described.  Two definitions, the maximum and offset
octahedral stresses, are then used to evaluate the yield stress for a
wide range of model parameters.  In all cases, the onset of shear is
consistent with the pressure-modified von Mises criterion, and the
pressure coefficient is nearly independent of many parameters.  Under
triaxial tensile loading, the mode of failure changes to cavitation.
\end{abstract}

\pacs{PACS numbers: 83.60.La, 62.20.Fe, 83.10.Mj}

\begin{multicols}{2}\narrowtext

\section{Introduction}\label{intro-sec}

The ability to predict the conditions under which a material will
yield is of great fundamental interest and technological
importance \cite{Haward1997,Perez1998,Ward1983,Courtney1990}.
Over the past three centuries, a number of yield criteria have been
formulated that predict whether a combination of stresses on a solid
will produce irreversible deformation.  In this paper, we use
molecular dynamics simulations to test the applicability of these
criteria to amorphous polymer glasses under multiaxial loading.

The yield criterion that is most commonly used for polymer glasses is
the pressure-modified von Mises (pmvM) criterion. The original von
Mises criterion \cite{vonMises1913} was based on the assumption that
yield occurs in the material when the elastic free energy associated
with the shear deformation $F_{\rm shear}$ reaches a critical value.
For small deformations of an isotropic material, $F_{\rm shear}$ is
proportional to the square of the octahedral or deviatoric stress
\begin{eqnarray}
\tau_{\rm oct}&=& \frac{1}{3}\left((\sigma_{1}-\sigma_{2})^2+(\sigma_{2}-\sigma_{3})^2+(\sigma_{3}-\sigma_{1})^2\right)^{1/2}\nonumber\\
&=&\frac{1}{\sqrt{3}}\left((\sigma_{1}+p)^2+(\sigma_{2}+p)^2+(\sigma_{3}+p)^2\right)^{1/2},
\end{eqnarray}
where the $\sigma_i$ denote the three principal stress components and
$p=-(\sigma_1+\sigma_2+\sigma_3)/3$ is the hydrostatic pressure. If
one ignores anharmonic effects, the condition of yield at constant
$F_{\rm shear}$ can be reformulated as yield at a constant threshold
value of $\tau_{\rm oct}$, i.e.
\begin{equation}
\label{vonmises-eq}
\tau_{\rm oct}^y=\tau_0.
\end{equation}

For polymers it is often observed \cite{Whitney1967} that tensile
pressure promotes yielding and compressive pressure delays yielding,
which is not accounted for in Eq.~(\ref{vonmises-eq}). The pressure
modified von Mises criterion \cite{Bauwens1970} includes a linear
pressure dependence in the value of the octahedral shear stress at
yield,
\begin{equation}
\label{pressure-eq}
\tau_{oct}^{y}=\tau_0+\alpha p.
\end{equation}
This pressure modification is motivated by an analogy to friction,
where the shear stress is linearly related to normal pressure, rather
than to an energy argument like that used to motivate
Eq.~(\ref{vonmises-eq}). The coefficient $\alpha$ then plays the role
of an internal friction coefficient. Eq.~(\ref{pressure-eq}) is also
employed to describe the shear yield stress in granular
materials \cite{deGennes1999}.

The pressure-modified Tresca criterion (pmT) is also often quoted in
the context of polymer yielding. Here the relevant quantity is assumed
to be the maximum shear stress 
\begin{equation}
\tau_{max}=\frac{1}{2}|\sigma_i-\sigma_j|_{\rm max},
\label{tmax-eq}
\end{equation}
and yield is assumed to occur at
\begin{equation}
\label{tresca-eq}
\tau_{max}^{y}=\frac{3}{\sqrt{2}}(\tau_0+\alpha p).
\end{equation}
The prefactor in Eq.~(\ref{tresca-eq}) is chosen so that, for the same
$\tau_0$ and $\alpha$, the pmT and pmvM criteria coincide when any two
of the three principal stress components are identical. In all other
cases, the pmT criterion gives a lower yield stress.

Several experimental studies are available that address the
macroscopic yield behavior of polymers. In these experiments, the
material is typically strained at a constant strain rate using
convenient geometries. The most commonly studied stress states are
uniaxial tension or compression, pure shear, and plane strain
compression. Several different definitions of the yield point have
been employed to determine the yield stress. Most authors
\cite{Bauwens1970,Bowden1972,Bubeck1984} take the maximum of the
stress-strain curve as the yield stress. The offset-stress
\cite{Raghava1973} is also used, particularly when the stress-strain
curve exhibits no clear maximum. It is given by the intersection of
the stress-strain curve with a straight line that has the same initial
slope as the stress-strain curve but is offset on the strain-axis by a
specified strain, e.g. 0.2\% (see Figure \ref{yieldonset-fig}). The
offset stress is motivated by the idea that if the load was removed,
the sample would relax elastically along the straight line to an
unrecoverable strain that is equal to the offset stress. However, the
relaxation is generally more complicated. The authors of
Ref.~\cite{Quinson1997} argue that the true yield point ought to be
determined from the residual strain measured after unloading. They
define the yield stress as the smallest stress value that gives a
nonzero residual strain.

Raghava et al.~\cite{Raghava1973} collected yield data for
polyvinylchloride (PVC), polystyrene (PS), polymethylmethacrylate
(PMMA), and polycarbonate (PC). Despite the fact that different
definitions of yield were used, all data fit the pmvM
criterion\cite{foot1} with the same value of $\alpha\simeq 0.18$. More
recently, Quinson et al.~\cite{Quinson1997} concluded that PMMA
is always well described by the pmvM criterion, but that PS follows
the pmT criterion at low temperatures and PC always follows the pmT
criterion. They associated the validity of the pmT criterion with the
occurrence of shear bands. However, Bubeck et
al.~\cite{Bubeck1984} measured a larger number of yield points
for PC and found that they fit the pmvM criterion. This discrepancy
may be due to the use of different yield point definitions in the two
studies. The small number of uni- and biaxial stress states considered
in experimental studies makes it hard to distinguish the two
criteria. No experimental tests of yield criteria appear to have
considered triaxial stress states.

On the simulation side, both Monte Carlo \cite{Chui1999} and
Molecular Dynamics simulations \cite{Yang1997} have been
presented for uniaxial deformation of polymer glasses.
Coarse-grained stochastic dynamics of shear flow of local volumes has
also been considered \cite{Argon1995}. These models were shown
to qualitatively reproduce experimental stress-strain curves and their
dependence on temperature and strain rate. However, we are unaware of
any studies that address multiaxial stress states and relate them to
yield criteria.

This paper is organized as follows. In Section \ref{model-sec} we
discuss the details of our model interaction potential for polymer
glasses and our method for imposing and measuring stress and
strain. Section \ref{results-sec} begins with a comparison of yield
stresses calculated from different definitions of the yield point.  We
then determine both the maximum stress and the offset stress in
multiaxial stress states and compare the results to the pmvM and pmT
criteria. In particular, we address the nature of the pressure
dependence of the yield stress and investigate the validity of
Eqs.~(\ref{pressure-eq}) and (\ref{tresca-eq}) in different physical
limits. We conclude with a discussion in Section \ref{discussion-sec}.

\section{Polymer Model and Simulations}\label{model-sec}
Our model of a polymer glass builds upon extensive simulations of
polymer melt dynamics \cite{Kremer1990}. The solid consists of
linear polymers each containing N beads. Pairs of beads of mass
$m$, separated by a distance $r$, interact through a truncated
Lennard-Jones (LJ) potential of the form
\begin{eqnarray}
V_{\rm LJ}(r)&=&4u_0\left[(d/r)^{12}-(d/r)^{6}\right]\qquad{\rm
for\,\,}r<r_c\\ \nonumber &=&0 \qquad{\rm for\,\,}r>r_c,
\end{eqnarray}
where $u_0$ and $d$ are characteristic energy and length scales.
Adjacent beads along the chain are coupled by the FENE potential
\cite{Kremer1990}
\begin{equation}
V_{\rm FENE}(r)=-\frac{1}{2}kR_0^2\ln[1-(r/R_0)^2]\qquad{\rm for\,\,}r<R_0,
\end{equation}
where $R_0=1.5\,d$ and $k=30\,u_0/d^2$ are canonical choices that
yield realistic melt dynamics \cite{breakablechaincomm}. Mappings of
the potential to typical hydrocarbons \cite{Kremer1990} give
values of $u_0$ between 25 and 45 meV and $d$ between 0.5 and 1.3 nm.

The above potentials produce very flexible polymer chains, since no
third-, or fourth-body potentials are present. Melt studies of the
bead-spring model have recently been extended to include the effect of
polymer rigidity \cite{Faller2000}. Following these authors, we
consider a bond-bending potential for each chain,
\begin{equation}
V_{B}=\sum_{i=2}^{N-1}b\left(1-\frac{(\vec{r}_{i-1}-\vec{r}_{i})\cdot(\vec{r}_{i}-\vec{r}_{i+1})}{|(\vec{r}_{i-1}-\vec{r}_{i})||(\vec{r}_{i}-\vec{r}_{i+1})|}\right),
\label{semiflex-pot}
\end{equation} 
where $\vec{r}_{i}$ denotes the position of the $i$th bead along the
chain, and $b$ characterizes the stiffness.  

The equations of motion are integrated using the velocity-Verlet
algorithm with a timestep of $dt=0.0075\,\tau_{\rm LJ}$, where
$\tau_{\rm LJ}=\sqrt{md^2/u_0}$ is the characteristic time given by
the LJ energy and length scales. Periodic boundary conditions are
employed in all directions to eliminate edge effects. We consider two
system sizes of 32768 beads and 262144 beads at two temperatures of
$T_{\rm h}=0.3\,u_0/k_B$ and $T_{\rm l}=0.01\,u_0/k_B$. $T_{\rm h}$ is
very close to the glass transition temperature $T_g\approx
0.35\,u_0/k_B$ for this model \cite{Bennemann1998}, and $T_{\rm
l}$ represents an effectively athermal system. The temperature is
controlled with a Nos\'e-Hoover thermostat (thermostat rate
$2\,\tau_{\rm LJ}^{-1}$) \cite{Hoover1985}.

\begin{figure}[hbt]
\epsfig{file=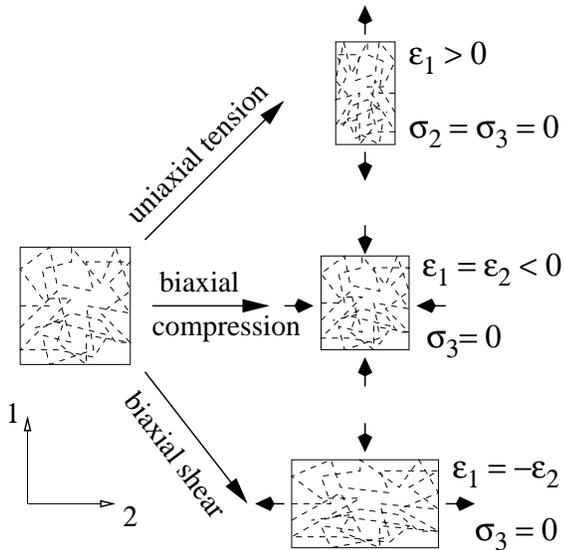,width=7.5cm}
\caption{Illustration of some limiting stress states employed in this
study. For uniaxial simulations $\sigma_2$ and $\sigma_3$ were
kept at zero using a Nos\'e-Hoover barostat.
For biaxial cases $\sigma_3$ was kept at zero.
The controlled strains were increased at a constant rate.
}
\label{strainillu-fig}
\end{figure}

To study the yield behavior, we impose tensile or compressive strains
$\epsilon_{i}$ on one or more axes of the initially cubically
symmetric solid at constant strain rates of $d\epsilon_i/dt=
10^{-4}\,\tau_{\rm LJ}^{-1}$ or less. Figure \ref{strainillu-fig}
illustrates some of the important limiting stress states that are
discussed below. For uni- and biaxial studies of yielding, the
stresses $\sigma_{i}$ in the remaining two or one directions are
maintained at zero by a Nos\'e-Hoover barostat (barostat rate
$0.1\,\tau_{\rm LJ}^{-1}$) \cite{Hoover1985}. Note that since
$\tau_{\rm LJ}\sim 3\,{\rm ps}$, the strain rates employed are 8-9
orders of magnitude higher than typical experimental strain
rates. Nevertheless by comparing to typical sound velocities we can
ensure that our strain rates are slow enough that stresses can
equilibrate across the system and thus loading proceeds nearly
quasistatically.  A detailed study of the rate dependence of yielding
in polymers will be presented elsewhere.
All stresses in this paper are true stresses, i.~e. $\sigma=F/A$,
where A is the instantaneous cross-sectional area. Likewise, we use
true or logarithmic strains
$\epsilon=\int_{l_0}^{l}\frac{dl}{l}=\ln(l/l_0)$. In all cases,
stresses are averaged over the entire simulation cell.

Our model lets us address a variety of different physical situations.
The influence of adhesive interactions on yielding can be examined by
varying the LJ potential cutoff distance from $r_c=2^{1/6}\,d$, which
gives a purely repulsive interaction, to $r_c=1.5\,d$ or $r_c=2.2\,d$,
which include progressively longer ranges of the attractive tail of
the LJ potential.  As a byproduct, the density of the solid at zero
pressure changes with $r_c$.

Polymer dynamics in the melt is greatly influenced by chain length,
particularly when entanglements are present. To study the effect of
chain length on yielding, we consider an entangled case $N=256\approx
4\times N_e$ and an unentangled case $N=16\approx 1/4\times N_e$,
where $N_e\approx 60$ is the entanglement length of the flexible
bead-spring model \cite{Puetz2000}. The bond-bending potential
$V_B$ is also varied from the flexible limit $(b=0\,u_0)$ to a
semiflexible case $(b=1.5\,u_0)$.

\section{Results}\label{results-sec}

\subsection{Onset of yielding and yield stress definitions}
\begin{figure}[hbt]
\epsfig{file=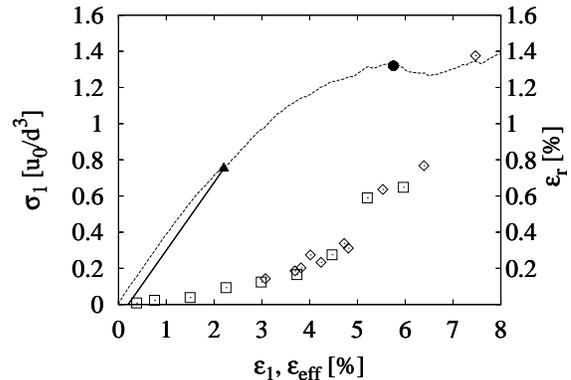,width=7.5cm}
\caption{Stress-strain curve (dashed line) in uniaxial loading. The
maximum stress is indicated by $\bullet$ and the 0.2\% offset stress
by $\blacktriangle$.  The latter is determined by the intersection
with the straight line, which has the same initial slope as the
stress-strain curve, but is displaced by 0.2\% on the strain axis.
Also shown is the residual strain $\epsilon_r$ as a function of the
maximum strain during uniaxial ($\square$) or biaxial ($\diamond$)
loading. The biaxial results are plotted against the effective strain
$\epsilon_{\rm eff}$ (Eq.~(\ref{effstrain-eq})). Here $T=T_{\rm l},
N=256, r_c=1.5\,d, b=0\,u_0,$ and there are 32768 beads.}
\label{yieldonset-fig}
\end{figure}

We begin by considering a uniaxial tension simulation, in which the
polymer glass is expanded in one direction, while zero stress is
maintained in the perpendicular plane. Figure \ref{yieldonset-fig}
shows the resulting stress-strain curve. It exhibits the typical
features of a polymer glass: linear response is followed by a
nonlinear region that bends over into a maximum. Note that the maximum
is reached at at strain of about 6\%, which is in very good agreement
with experimental stress strain curves, see
e.~g.~\cite{Quinson1997}. Once past that maximum, the stress
drops slightly and the material begins to flow, but these large
strains are not shown.
Fig.~\ref{yieldonset-fig} also illustrates how to obtain the maximum
and offset yield stresses as described in the introduction. Both
quantities are easy to identify for uniaxial stress. To make contact
with the approach of ref.~\cite{Quinson1997}, we have also
determined the amount of residual strain after unloading. Squares show
the residual strain $\epsilon_r$ as a function of the maximum strain
before unloading. Results for $\epsilon_r>0.2\%$ are comparable to the
residual strains given in ref.~\cite{Quinson1997}. These authors
identified the yield point with the strain where a linear
extrapolation of $\epsilon_r$ vanishes. We have extended our studies
to much smaller residual strains and find no evidence of a threshold
strain below which the residual strain vanishes. Studies of other
glassy systems also suggest that at $T=0$ a finite unrecoverable
strain should be present for arbitrarily small initial strains, but
the theoretical situation at finite $T$ is more
complicated \cite{gruneretc}.

Note that in order to obtain 0.2\% residual strain, the initial
extension has to be about 4\%. This is about twice as large as the
yield strain from the 0.2\% offset criterion. The offset criterion
assumes elastic recovery along the line determined by the elastic
moduli of the unperturbed solid. Examination of the recovery curves
shows that they are non-linear. This confirms that much of the
nonlinear response under loading to the triangle in
Fig.~\ref{yieldonset-fig} is due to recoverable, but anharmonic
deformations.

The above discussion raises questions about both the applicability of
the unrecoverable strain definition of the yield point and
the motivation for the offset stress definition.
However, since the offset stress is widely used in engineering
applications, we consider both it and the maximum stress definition in
the following sections.

\subsection{Evaluating the yield stress in multiaxial stress states}\label{yieldstressdef-subsec}

\begin{figure}[hbt]
\epsfig{file=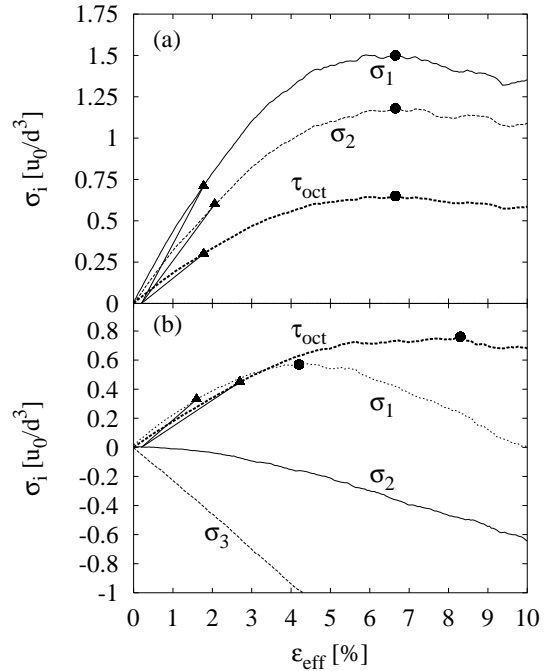,width=7.5cm}
\caption{Nonzero principal stress components in the model polymer
glass during (a) biaxial and (b) triaxial loading. The maximum and
offset yield stresses are indicated by $\bullet$ and $\blacktriangle$,
respectively. Neither the maximum nor the offset definition can be
applied to two of the three components in (b). Also shown is the
octahedral shear stress $\tau_{\rm oct}$ (thick dashed line) and the
yield stresses that result by applying the yield point definitions to
it. Here $T=T_{\rm l}, N=256, r_c=1.5\,d, b=0$, and there are 32768
beads.}
\label{yieldstressdef-fig}
\end{figure}

Figure \ref{yieldstressdef-fig} shows the principal stress components
$\sigma_i$ as well as the octahedral shear stress $\tau_{\rm oct}$ for
(a) biaxial and (b) triaxial loading. In multiaxial stress states one
needs to plot the stress(es) against a scalar effective strain
$\epsilon_{\rm eff}$ constructed from the three principal components
$\epsilon_i$ of the strain tensor. This choice is not unique, and
influences the value obtained for the offset stress. A convenient
choice that we employ in this paper is
\begin{equation}
\label{effstrain-eq}
\epsilon_{\rm eff}=\frac{1}{\sqrt{2}(1+\nu)}\left((\epsilon_{1}-\epsilon_{2})^2+(\epsilon_{2}-\epsilon_{3})^2+(\epsilon_{3}-\epsilon_{1})^2\right)^{1/2},
\end{equation}
where $\nu$ is Poisson's ratio. This expression is proportional to the
octahedral strain and reduces to $\epsilon_1$ under uniaxial
loading. It also reduces to the effective strain defined in
Ref.~\cite{Raghava1973} for biaxial strains, and allows us to
treat all strain states on the same footing.

Fig.~\ref{yieldstressdef-fig}(a) illustrates the maximum and offset
definitions of the yield point for biaxial stresses. The maximum of
the stress-strain curves is straightforward to evaluate (solid
circles), and both nonzero stress components peak at the same
strain. By contrast, the offset definition gives yield at slightly
different strains (solid triangles). This is troubling, since a yield
point should refer to a unique instance in time or strain.  Moreover,
the offset stress depends on arbitrary factors in the definition of
the effective strain and in the amount of offset at the yield point.

We also evaluated the residual strain for different biaxial loading
conditions and added them to Fig.~\ref{yieldonset-fig}
(diamonds). Interestingly, these residual strains fall onto the same
curve as the results from uniaxial extension when plotted against
$\epsilon_{\rm eff}$. As above, there is no indication of a finite
threshold below which the residual strain vanishes.

\begin{figure}[hbt]
\epsfig{file=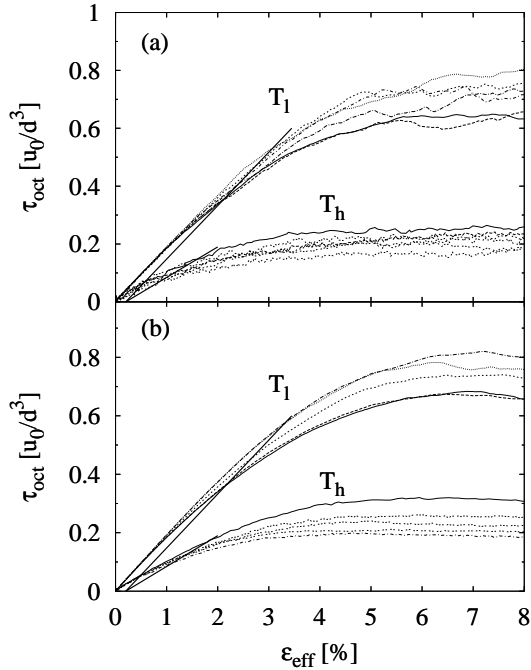,width=7.5cm}
\caption{Octahedral shear stress $\tau_{oct}$ versus effective strain
in simulations with (a) 32768 and (b) 262144 beads at
$T_{\rm h}=0.3u_0/k_B$ and $T_{\rm l}=0.01u_0/k_B$ for different
stress states. At low strains, all curves
for a given temperature collapse onto a line whose slope is determined
by the elastic moduli (Eq.~(\ref{elmod-eq})).
At higher strains the
curves split apart, with the higher curves corresponding to higher
pressures.
Straight lines are drawn at
an offset of 0.2\% strain for each temperature. Here $N=256,
r_c=1.5\,d$, and $b=0\, u_0$.}
\label{stressstrain15-fig}
\end{figure}

Fig.~\ref{yieldstressdef-fig}(b) shows that determining the yield
point is even more complicated under triaxial loading. For this case,
only the tensile stress component exhibits a clear maximum. The
compressive components decrease monotonically over the range
shown. Moreover, the slope of these curves increases in magnitude with
strain so that they do not intersect the line for the offset
criterion.
In order to make progress for general multiaxial stress states, we
thus decided to apply both the maximum and offset stress definitions
not to the three stress components individually, but to the octahedral
shear stress $\tau_{\rm oct}$. This is a natural choice given that
$\tau_{\rm oct}$ is the quantity that enters the pmvM criterion. The
method is illustrated in Fig.~\ref{yieldstressdef-fig} and again in
Fig.~\ref{stressstrain15-fig}, where $\tau_{\rm oct}$ is plotted
against the effective strain in a variety of uni-, bi-, and triaxial
stress states. For the biaxial states considered in experiments the
yield strains obtained from $\sigma_1$ and $\sigma_2$ separately are
nearly the same as that determined from $\tau_{\rm oct}$
\cite{Raghava1973}. Thus the value of $\tau_{\rm oct}^y$ is the
same as that constructed from the separate yield components.

In the initial elastic response, stress and strain tensors are
linearly related by a combination of the elastic moduli $c_{ij}$ of
the solid. The moduli are defined by $c_{ij}\epsilon_i=\sigma_j$, from
which we obtain
\begin{equation}
\label{elmod-eq}
\tau_{\rm oct}=\frac{\sqrt{2}}{3}\frac{(c_{11}+2c_{12})(c_{11}-c_{12})}{c_{11}+c_{12}}\epsilon_{\rm
eff}
\end{equation}
by straightforward insertion. Thus a plot of $\tau_{\rm oct}$ versus
$\epsilon_{\rm eff}$ should collapse all curves onto a stress-strain
curve with common initial slope. Fig.~\ref{stressstrain15-fig} shows
that this condition is satisfied for two different temperatures and
system sizes. All curves start out with the same slope, but they split
apart rapidly at higher strains. This reflects the influence of
pressure on yielding. We will show later that this pressure dependence
is well described by Eq.~(\ref{pressure-eq}) for both yield point
definitions. As in experiments, the yield stresses decrease with
increasing temperature.
The larger systems in Fig.~\ref{stressstrain15-fig}(b) exhibit much
smoother curves, since many more independent yield events contribute
to the average stress response. However, since the values of yield
stress and strain do not change significantly with size, we use
systems with 32768 beads in most cases. This speeds up the
computations and allows a larger parameter space to be explored. Note
that each curve represents the stress response for one particular
initial configuration of polymers in the glassy state. We have
observed variations on the order of 5\% in the magnitude of $\tau_y$
with different 32768 bead configurations, but the general features of
yielding remained unaltered.

\subsection{Biaxial yielding}
Experimental tests of yield criteria have generally considered
uniaxial or biaxial loading conditions. In the biaxial case, the pmvM
criterion predicts that the values of $\sigma_1$ and $\sigma_2$ at
yield should lie on the surface of an ellipse in the
$\sigma_1-\sigma_2$ - plane. The equation for this ellipse can be
obtained from Eq.~(\ref{pressure-eq}) by letting $\sigma_3=0$:
\begin{eqnarray}
\label{ellipse-eq}
(\sigma_1-\sigma_2)^2&+&\frac{1-2\alpha}{3}\left(\sigma_1+\sigma_2+\frac{3\alpha\tau_0}{1-2\alpha}\right)^2\nonumber\\
&=&6\tau_0^2\left(1+\frac{\alpha^2}{2(1-2\alpha)}\right)
\end{eqnarray}

A nonzero value of $\alpha$ breaks the tensile/compressive symmetry
and shifts the ellipse towards the lower left quadrant (where the
stress is compressive). The amount of the shift increases with
$\alpha$.  The pmT criterion predicts a yield surface bounded by
straight lines. For the same $\tau_0$ and $\alpha$, it coincides with
the pmvM criterion whenever two of the stresses are equal.
\begin{figure}[hbt]
\epsfig{file=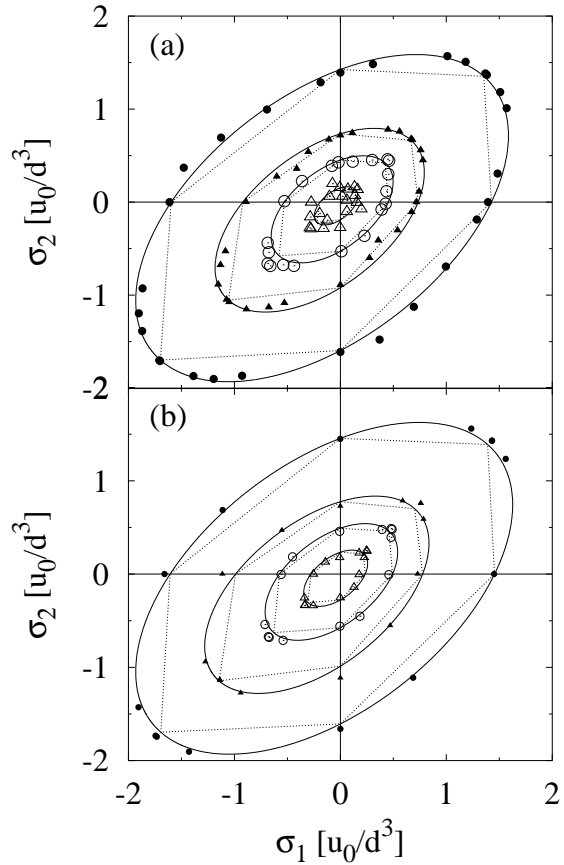,width=7.5cm}
\caption{Yield points for biaxial stress states. Filled symbols
$(\bullet,\blacktriangle)$ correspond to the yield point at $T_{\rm
l}$ and open symbols $(\circ,\triangle)$ to $T_{\rm h}$. Circles refer
to the maximum yield stress and triangles to the offset-stress. Data
in (a) was obtained for 32768 and in (b) for 262144 beads. The
ellipses represent the pmvM yield surface, Eq.~(\ref{ellipse-eq}),
and the dashed lines show
the pmT yield surface. Statistical error bars are indicated by the
symbol size. Here $N=256, r_c=1.5\,d$, and $b=0\,u_0$.}
\label{biaxial-fig}
\end{figure}
In Fig.~\ref{biaxial-fig}, we perform an analysis analogous to the
experimental studies, where we only include biaxial data from
Fig.~\ref{stressstrain15-fig}. We employ reflection symmetry about the
line $\sigma_1=\sigma_2$, since the solid is isotropic. We have
tested this isotropy in selected cases and find variations on the
order of the symbol sizes. The offset stress is generally lower than
the maximum stress and becomes very difficult to evaluate at $T_{\rm
h}$ due to the small magnitude of the yield stress and the large thermal
noise. The yield stresses are larger in the lower left quadrant,
indicating that yield is pressure-dependent. Also shown are the yield
surfaces predicted by the two yield criteria. As can be seen, the data
from both definitions of the yield stress are better described by the
pmvM criterion (Eq.~(\ref{ellipse-eq})) than the pmT criterion. The
values of $\tau_0$ and $\alpha$ can be found in Table I (see also the
next section). In the following discussion we focus on the pmvM
criterion, which provides a better fit for all cases studied.

\subsection{General stress states} \label{genstress-subsec}
\begin{figure}[hbt]

\epsfig{file=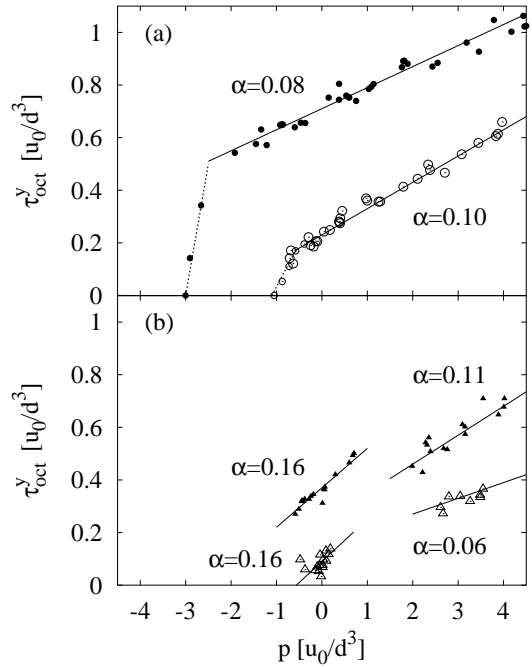,width=7.5cm}
\vspace*{0.2cm}
\caption{Octahedral stress at yield versus pressure at $T_{\rm l}$
(filled symbols) and $T_{\rm h}$ (open symbols). The maximum stress is
shown in (a) and the offset stress in (b). Solid lines are fits to
Eq.~(\ref{pressure-eq}), and the values for $\tau_0$ and $\alpha$ are
given in Table I. Results for $p>1.5\,u_0/d^3$ were obtained by
applying strains to a compressed state (see text). Points where
cavitation was observed are connected by a dashed line in (a) and are
not shown in (b).}
\label{taup-fig}
\end{figure}

For general stress states it is simplest to test
Eq.~(\ref{pressure-eq}) directly by plotting $\tau_{oct}^y$ versus
pressure. This plot lets us include triaxial stress states, which
could not be discussed in Fig.~\ref{biaxial-fig}. In
Fig.~\ref{taup-fig}(a) we show the maximum shear yield stresses from
all stress states at two temperatures. In general one would expect
Eq.~(\ref{pressure-eq}) to hold as long as the mode of failure is
shear. The linear law should break down when cavitation occurs in the
polymer. Dashed lines in Fig.~\ref{taup-fig}(a) indicate the
threshold for cavitation, which only occurs in our model when all
three stress components are tensile. For isotropic tensile loading
$\tau_{\rm oct}^y$ is rigorously zero and can not be used to determine
a maximum or offset stress.  The endpoint at $\tau_{\rm oct}^y =0$ in
Fig.~\ref{taup-fig}(a) was determined from the pressure where all three
stress components reach a simultaneous maximum.  The offset stress is
not well-defined in this limit because $\epsilon_{\rm eff}$ also
vanishes.

When yield occurs through shear, the yield stresses shown in Fig.
\ref{taup-fig}(a) lie on a straight line as predicted by
Eq. (\ref{pressure-eq}).  To test this linear dependence over a wider
range, we have also performed simulations in which the simulation cell
was first compressed isotropically to a higher pressure. The same
range of strains was then imposed again, and the resulting shear
stresses were added to Fig.~\ref{taup-fig}(a). This procedure led to
all the points for $p>1.5\,u_0/d^{3}$, which lie on the same straight
line as data from the uncompressed starting state. Other simulations
show that this linearity extends to even greater pressures of up to
$20 u_0/d^3$.

Fig.~\ref{taup-fig}(b) shows the offset yield stresses for two
temperatures.  While, the data from the compressed and uncompressed
starting states each fall on a straight line, it is no longer possible
to make a common fit through both data sets.  
Moreover, the values of $\alpha$ tend to be smaller for the compressed
state than the uncompressed state.  From Fig.~\ref{stressstrain15-fig}
one sees that the value of $\alpha$ from the offset criterion depends
on the rate at which stress increases with pressure at fixed strain,
and the slope of the stress-strain curve near the intersection.  Both
factors change with temperature and the choice of offset at yield.
For this reason, it is not surprising that values of $\alpha$ from the
offset criterion show larger variations than those from determined
from the maximum stress.

\subsection{Effect of adhesive interactions}
\label{longercutoff-subsec}
Intuitively one would expect that increasing adhesive interactions
between polymers would lead to a higher yield stress. In our model
this can be achieved by increasing $r_c$, which controls how much of
the attractive tail in the LJ potential is included.  We have verified
that increasing $r_c$ to $2.2\,d$ produces higher yield stresses and
that the yield points are consistent with the pmvM criterion.  Biaxial
data lie on an ellipse rather than the straight line segments
predicted by the pmT criterion (Fig. \ref{biaxial-fig}). Values of
$\tau_{\rm oct}^y$ from general stress states fall onto a straight
line when plotted against pressure (see Fig.~\ref{collapse-fig} below)
as long as failure occurs through shear rather than cavitation.  Fit
values of $\tau_0$ and $\alpha$ are presented in Table I. The value of
$\alpha$ is the same as for smaller $r_c$ within our errorbars. The
value of $\tau_0$ increases, leading to a higher yield stress at all
pressures.  Increasing $r_c$ also suppresses cavitation. The magnitude
of the tensile pressure needed to produce cavitation under hydrostatic
loading increased by 2 $u_0/d^{3}$ at both temperatures.
\begin{figure}[hbt]
\epsfig{file=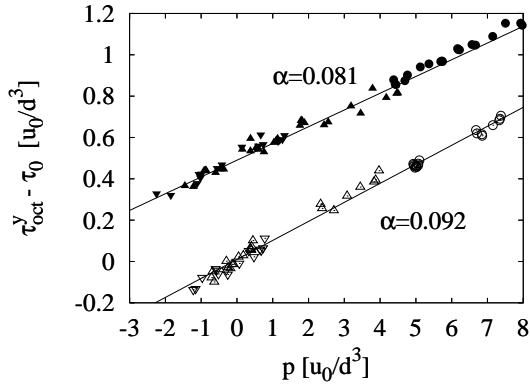,width=7.5cm}
\vspace*{0.2cm}
\caption{Maximum shear stress $\tau_{\rm oct}^y$ minus $\tau_0$ at
$T_{\rm l}$ (filled symbols) and $T_{\rm h}$ (open symbols) for three
different potential cutoffs $r_c=1.1225\, d$ $(\bullet,\circ)$,
$r_c=1.5\, d$ $(\blacktriangle,\triangle)$ and $r_c=2.2\, d$
$(\blacktriangledown,\triangledown)$. To avoid overlap, results for
$T_{\rm l}$ were displaced vertically by $0.5\,u_0/d^3$. Points where
cavitation occurred are not included in this plot. Solid lines show
best fits to all points plotted.}
\label{collapse-fig}
\end{figure}
The opposite limit that we can consider is the purely repulsive
potential obtained when $r_c=2^{1/6}\, d$. Such a solid does not
provide any resistance to tensile stresses at $p=0$, but the same
range of strains considered above can be applied at higher offset
pressures as done in Fig.~\ref{taup-fig}. Once again, the results are
in good agreement with the pmvM criterion and values of $\tau_0$ and
$\alpha$ are quoted in Table I. The value of $\tau_0$ is greatly
reduced from those for larger $r_c$, but $\alpha$ remains unchanged
within our errorbars.

To illustrate that $\alpha$ is constant, we summarize data for the
maximum shear yield stress as a function of pressure for all three
different ranges of the potential in Fig.~\ref{collapse-fig}.  For
each case, the value of $\tau_0$ is subtracted, and results for
$T_{\rm l}$ are displaced vertically by 0.5 $u_0/d^{3}$ in order to
avoid overlap with the high temperature data.  For each temperature,
all of the data points collapse onto a common straight line whose
slope is consistent with the independently determined values of
$\alpha$.  Thus while $\tau_0$ decreases rapidly as adhesion is
reduced, $\alpha$ remains unchanged.  Similar results have been
obtained in recent studies \cite{He} of static friction as summarized
in Sec.~\ref{discussion-sec}.

\subsection{Semiflexible and short polymers}\label{semiflexible-subsec}
In the previous section we showed that the interaction range has a
large effect on $\tau_0$, but does not significantly alter
$\alpha$. Other features of the potential have little effect
on either quantity.

\begin{figure}[hbt]
\epsfig{file=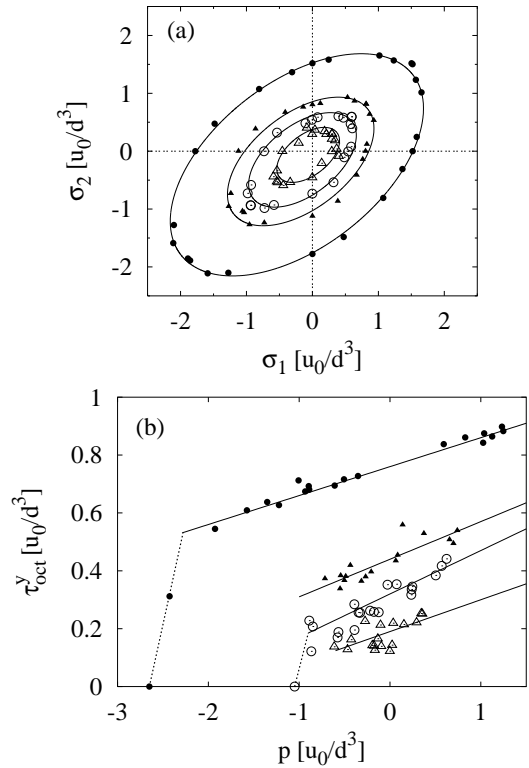,width=7.5cm}
\vspace*{0.2cm}
\caption{Yield points for (a) biaxial stress states and (b) general
stress states from simulations with semiflexible chains $(b=1.5\,u_0)$. As before,
filled (open) symbols correspond to $T_{\rm l}$ ($T_{\rm h}$) and
circles (triangles) indicate the maximum (offset) yield stress.  Solid
lines show fits to the pmvM criterion using values of $\alpha$ and
$\tau_ 0$ given in Table I. Dashed lines connect points where
cavitation occurred. Here $N=256$ and $r_c=1.5\,d$ and there
are 32768 beads.}
\label{semiflex-fig}
\end{figure}

We first consider semiflexible polymers with a value of $b=1.5\,u_0$
and a cutoff $r_c=1.5\,d$ in Eq.~(\ref{semiflex-pot}). This makes the
polymer conformation more rigid, and one would therefore expect
increased resistance to deformation. Fig.~\ref{semiflex-fig} (a) shows
that the pmvM criterion also describes yield of this semiflexible
polymer. Values of $\tau_0$ and $\alpha$ from linear fits in
Fig.~\ref{semiflex-fig}(b) tend to be slightly higher than the values
for flexible chains (Table I), but the changes are comparable to
statistical uncertainties.

Finally, we have considered shorter chains of length $N=16$ to
contrast the behavior of unentangled and entangled ($N=256$) chains.
The long-range topological properties that are crucial for the
dynamics of polymer melts turn out to be irrelevant for the small
strains required to initiate yield \cite{foot2}.  Values for $\tau_0$
and $\alpha$ do not deviate substantially from the results for
$N=256$, and we merely quote them in Table I.

\section{Summary and Discussion}\label{discussion-sec}
We have examined the yield stress of amorphous glassy polymers under
multiaxial loading conditions using molecular dynamics simulations.
Measurements of the residual strain after unloading indicate that
imposing any finite strain produces some permanent plastic
deformation.  Thus no threshold for the onset of plastic yield could
be identified \cite{Quinson1997}.  Two other common definitions
of the yield point, the maximum and offset stress, could not be
applied separately to the three principal stresses.  Instead they were
applied to the octahedral shear stress $\tau_{\rm oct}$ for a wide
range of loading conditions and model parameters.

In all cases studied, and for both yield definitions, the
pressure-modified von Mises criterion provides an adequate description
of the shear yield stress. Traditional plots of biaxial yield points
are well fit by ellipses (Eq.~(\ref{ellipse-eq})) and multi-axial
results for $\tau_{\rm oct}^y$ vary linearly with pressure. The
pressure-modified Tresca criterion always shows worse agreement.
Although the difference may be small from an engineering perspective
($<21$\%), it has implications about the nature of the molecular
mechanisms that lead to yield.

More experimental studies seem to be consistent with the pmvM
criterion than the pmT criterion, although the number of stress states
that has been accessed is relatively small and limited to biaxial
cases.  One study \cite{Quinson1997} found that the pmT
criterion worked better for polymers that formed shear bands.  Shear
banding can be influenced by boundary conditions and the method of
imposing shear.  In particular, our periodic boundary conditions
inhibit shear banding and this may be why the pmvM criterion always
provided a better fit to the yield stress.  It is also possible that
the failure criterion is dependent on the specific form of the atomic
interactions, and that the simple potentials considered here do not
lead to pmT behavior.  These will be fruitful subjects for future
research.

The dependence of $\tau_0$ and $\alpha$ on potential parameters and
temperature was explored.  The value of $\alpha$ is nearly independent
of potential parameters, but seems to increase slightly with
increasing temperature. The value from the offset stress is larger and
more variable than the value from the maximum stress.  The value of
$\tau_0$ is also relatively insensitive to the chain length $N$ and
degree of rigidity, indicating that yield is dominated by local
structure. However, $\tau_0$ increases rapidly with increases in the
attractive interactions between molecules that determine the cohesive
energy.  The value of $\tau_0$ also decreases with increasing
temperature. Temperature and rate dependence will be discussed in
subsequent work.

Our calculated values of the dimensionless parameter $\alpha$ compare
well with typical experimental values.  Quinson et
al.~\cite{Quinson1997} report values for $\alpha$ between 0.14--0.25
for PMMA, 0.03--0.12 for PC and 0.18--0.39 for PS. As in our
simulations, there was a tendency for the value to increase with
temperature. Bubeck et al.~\cite{Bubeck1984} found values between
0.04--0.07 for PC.  Rhagava et al. fit all data for the above polymers
and PVC to $\alpha=0.18$, and best fit values for individual polymers
ranged from 0.13--0.20.  They used the offset stress to determine
$\alpha$ and our numbers for this quantity are in the same range:
$\alpha=$ 0.11--0.16.  Comparing values of $\tau_0$ is more
complicated because of uncertainty in mapping our potential parameters
to specific polymers (see Section \ref{model-sec}).  Using values of
$u_0$ and $d$ from Ref.~\cite{Kremer1990} our values of $\tau_0$
correspond to 1 to 50 MPa.  This coincides well with the experimental
range \cite{Ward1983,Quinson1997} of 5--100 MPa, but quantitative
comparisons for specific polymers are not possible without more
detailed models of atomic interactions.

It is interesting to note a connection between the results reported
here and recent studies of the molecular origins of static friction
\cite{nanotrib,He}.  Macroscopic measurements of the force needed to
initiate sliding can be understood as arising from a local yield
stress for interfacial shear that rises linearly with normal pressure
exactly as in the pmvM criterion (Eq.~\ref{pressure-eq}).  Moreover,
the results presented here parallel results from simulations of the
static friction due to thin layers of short chain molecules that are
present on any surface exposed to air \cite{He}.  These authors find
$\tau_0$ represents an increase in the effective normal pressure due
to adhesive interactions.  As in our study, $\tau_0$ increases with
the range of interactions, $r_c$, and the value of $\alpha$ is
insensitive to many details of the interactions.  A simple geometrical
model for $\alpha$ was developed based on the idea that surfaces must
lift up over each other to allow interfacial sliding.  It will be
interesting to determine whether a similar geometric model can be
developed for yield of bulk polymers.

\acknowledgements Financial support from the Semiconductor Research
Corporation (SRC) and the NSF through grant No. DMR-0083286 is gratefully
acknowledged. We also thank the Intel Corp.~for a donation of
workstations. The simulations were performed with LAMMPS
\cite{Lammps1999}, a classical molecular dynamics package developed
by Sandia National Laboratories.

\begin{table}
\caption{Values for $\tau_0$ and $\alpha$ obtained from the linear
fits in Figures \ref{taup-fig}, \ref{semiflex-fig} as well as the
simulations with $r_c=2.2\,d$. Unless otherwise noted, $N=256$ and
$b=0$. Typical statistical errors in fits for the maximum yield stress
are $\pm 0.02\,u_0/d^3$ for $\tau_0$ and $\pm 0.01$ for
$\alpha$. Errors are about three times larger for $\tau_0$ in the
purely repulsive case, and for all offset stress results.}

\vspace*{0.1cm}
\begin{tabular}{ccccc}
  & $T_{\rm l}$ max & $T_{\rm l}$ offset & $T_{\rm h}$ max & $T_{\rm h}$ offset\\
  & $\tau_0 / \alpha$ & $\tau_0/\alpha$ & $\tau_0/\alpha$ & $\tau_0/\alpha$\\
\tableline
$r_c=2^{1/6}\,d$ & $0.06 / 0.08 $& --- & $-0.42/0.09 $& ---\\
$r_c=1.5\,d$ & $0.72/ 0.08 $& $0.37/ 0.15 $& $0.23/ 0.10 $& $0.11/ 0.16$\\
$r_c=1.5\,d$, comp. & $0.72/ 0.08 $& $0.24/ 0.11 $& $0.23/ 0.10 $& $0.15/ 0.06$\\
$r_c=2.2\,d$ & $0.83/0.09$ & $0.53/0.16$ & $0.45/0.10$ & $0.28/0.09$\\
$r_c=1.5\,d, b=1.5$ & $0.76/0.10$ & $0.44/0.13$ & $0.31/0.15$ & $0.19/0.11$ \\
$r_c=1.5\,d, N=16$ & $0.68/0.07$ & $0.36/0.14$ & $0.20/0.12$  & $0.07/0.06$ \\
\end{tabular}
\end{table}

\end{multicols}

\begin{references}
\bibitem{breakablechaincomm} The FENE-potential creates unbreakable
chains, but the deformations in this study do not reach stresses that
would accomplish chain scission.
\end{references}
\end{document}